%% file: MechanicalNeuralNetwork.tex
\documentclass[aps,twocolumn, superscriptaddress]{revtex4-1}

\usepackage[usenames]{color}
\usepackage{xcolor}
\usepackage{upgreek}
\usepackage{amsmath}
\usepackage{amssymb} 
\usepackage{graphicx}
\usepackage[utf8]{inputenc}
\usepackage{physics}
\usepackage{empheq}
\usepackage{braket}
\usepackage{siunitx}
\usepackage{float}
\usepackage{tcolorbox}
\usepackage{enumitem}
\usepackage{placeins}
\usepackage{xcolor}
\usepackage[T1]{fontenc}

\renewcommand{\topfraction}{1.9}
\renewcommand{\bottomfraction}{1.9}
\renewcommand{\textfraction}{.00}
\renewcommand{\floatpagefraction}{2.0}
\renewcommand{\dbltopfraction}{1.0}
\renewcommand{\dblfloatpagefraction}{1.0}

\begin{document}
\title{Binary classification of spoken words with passive phononic metamaterials} 
\author{Tena Dub\v{c}ek}
\affiliation{Institute for Theoretical Physics, ETH Zürich, 8093 Zürich, Switzerland}
\affiliation{Institute for Clinical Neurophysiology, Clinic Lengg, 8008 Zürich, Switzerland}
\affiliation{Department of Health Sciences and Technology, ETH Zürich, 8092 Zürich, Switzerland}
\author{Daniel Moreno-Garcia}
\affiliation{School of Engineering, EPFL, 1015 Lausanne, Switzerland}
\author{Thomas Haag}
\affiliation{Institute for Geophysics, ETH Zürich, 8092 Zürich, Switzerland}
\author{Parisa Omidvar}
\affiliation{AMOLF, Science Park 104, 1098 XG Amsterdam, the Netherlands}
\author{Henrik R. Thomsen}
\affiliation{Institute for Geophysics, ETH Zürich, 8092 Zürich, Switzerland}
\author{Theodor S. Becker}
\affiliation{Institute for Geophysics, ETH Zürich, 8092 Zürich, Switzerland}
\author{Lars Gebraad}
\author{Christoph B\"arlocher}
\affiliation{Institute for Geophysics, ETH Zürich, 8092 Zürich, Switzerland}
\author{Fredrik Andersson}
\affiliation{Institute for Geophysics, ETH Zürich, 8092 Zürich, Switzerland}    
\author{Sebastian D. Huber}
\affiliation{Institute for Theoretical Physics, ETH Zürich, 8093 Zürich, Switzerland}
\author{Dirk-Jan van Manen}
\affiliation{Institute for Geophysics, ETH Zürich, 8092 Zürich, Switzerland}
\author{Luis Guillermo Villanueva}
\affiliation{School of Engineering, EPFL, 1015 Lausanne, Switzerland}
\author{Johan O.A. Robertsson}
\affiliation{Institute for Geophysics, ETH Zürich, 8092 Zürich, Switzerland}
\author{Marc Serra-Garcia}
\affiliation{Institute for Geophysics, ETH Zürich, 8092 Zürich, Switzerland}
\affiliation{AMOLF, Science Park 104, 1098 XG Amsterdam, the Netherlands}

\date{\today}

\input{0_abstract}

\maketitle

\input{1_introduction}

\input{2_design}
\input{3_realization}

\input{4_outlook}

\section*{Acknowledgements}
We thank Matija Varga, Miguel Moleron-Bermudez, Sander Tans, Martin van Hecke and Rafael Polania for enlightening discussions. We are grateful to Sven Friedel at COMSOL multiphysics for help with the meshing of our FEM models. This work was supported by the European Research Council (ERC) under the European Union’s Horizon 2020 Research and Innovation Programme (Grant Agreements No. 694407 and No. 771503) and under the Horizon Europe Programme (Grant agreement No. 101040117).
\section*{}
Correspondence can be addressed to Tena Dub\v{c}ek (dubcekt@ethz.ch) and Marc Serra Garcia (m.serragarcia@amolf.nl). The datasets generated and/or analysed during the current study are available from the corresponding authors on reasonable request. Codes used to design the sample are provided as supplementary materials.

\bibliography{bib}

\newpage

\appendix
\input{5_appendix}

\end{document}

%% file: 0_abstract.tex
\begin{abstract}

Mitigating the energy requirements of artificial intelligence requires novel physical substrates for computation. Phononic metamaterials have a vanishingly low power dissipation and hence are a prime candidate for green, always-on computers. However, their use in machine learning applications has not been explored due to the complexity of their design process: Current phononic metamaterials are restricted to simple geometries (e.g. periodic, tapered), and hence do not possess sufficient expressivity to encode machine learning tasks. We design and fabricate a non-periodic phononic metamaterial, directly from data samples, that can distinguish between pairs of spoken words in the presence of a simple readout nonlinearity; hence demonstrating that phononic metamaterials are a viable avenue towards zero-power smart devices. 

\end{abstract}

%% file: 1_introduction.tex
% \section{Introduction}
The success of deep learning models is based on encoding complex tasks as a combination of large linear transformations and nonlinear activation functions. A variety of technologies, from photonics~\cite{Ashtiani2022} to memristor crossbar arrays~\cite{Yao2020}, have been postulated to minimize the energy costs associated to these large linear transformations. Phononic resonators have energy losses that improve on linear passive electronic systems by several orders of magnitude. This is reflected in their quality factors, which quantify the number of periods that oscillations take to decay. Quality factors of several thousands are common in phononic resonators~\cite{Serra-Garcia2018}, and can reach billions in specifically optimised devices~\cite{doi:10.1126/science.aar6939, Beccari2022}, but are in the tens for electronic circuits~\cite{Helbig2020}. This near-dissipationlessness, combined with the capability of directly processing mechanical signals such as spoken commands without first transducing them into an electronic or photonic domain, make phononic resonators a prime candidate for zero-power edge computing applications. Although these striking advantages have been recognised in the context of both classical~\cite{doi:10.1126/science.1144793, doi:10.1126/sciadv.aay6946} and quantum~\cite{Wollack2022, vonLupke2022} computing, and the environmental impact of artificial intelligence is increasingly in the spotlight~\cite{Dhar2020}, phononic implementations of machine learning models remain largely unexplored. 

\begin{figure}[ht!]
	\centering
    \includegraphics[width=0.95\columnwidth]{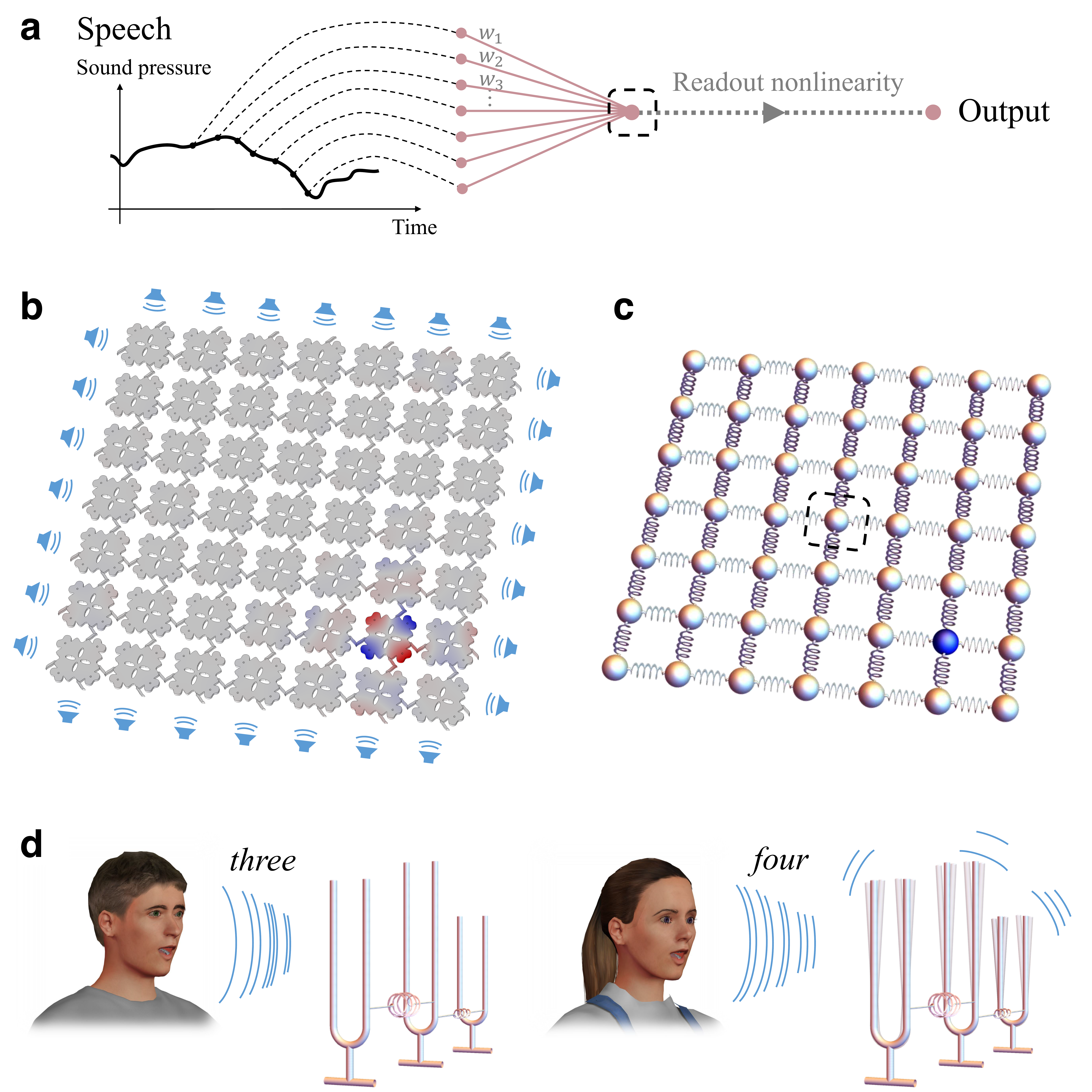}
\caption{ Passive speech recognition | 
{\fontfamily{phv}\selectfont\textbf{a}}
Speech classification by a temporal convolutional network that combines delayed copies of the signal according to a set of weights and then applies a readout nonlinearity.
% A resonator discriminates between frequencies.
% Our metamaterial of coupled resonators discriminates between spoken digits.
% The metamaterial is a lattice of vibrating plates connected by beams, with a geometry (beam locations and hole sizes) optimized to achieve the desired selective response.
% The structure is modelled as a mass-spring model. Each mass corresponds to a vibration localised around a plate. The blue mass corresponds to the displacement represented by the coloring in panel c.
{\fontfamily{phv}\selectfont\textbf{b}}
We realize a passive instance of such a network by a lattice metamaterial, whose vibrating plates (resonators) are connected by beams. Its geometry (beam locations and hole sizes) is optimized to achieve the desired selective response.
{\fontfamily{phv}\selectfont\textbf{c}}
The structure is modelled as a mass-spring model. Each mass corresponds to a vibration localised at a particular plate. The blue mass corresponds to the displacement represented by the coloring in panel b.
{\fontfamily{phv}\selectfont\textbf{d}}
The optimized metamaterial can be interpreted as a network of coupled resonators that discriminates between two spoken digits.
}
\label{Fig1}
\end{figure}

In this paper we report on the experimental realisation of a passive phononic metamaterial for speech processing tasks, that implements a convolutional layer on the sound signal~[Fig.~\ref{Fig1}a]. We start by experimentally showing that, for linearly separable word pairs, a single-layer metamaterial can solve the binary classification problem with significant accuracy (higher than 90\% in most cases). We then theoretically demonstrate that, for words that are not linearly separable, we can achieve a good classification performance by constructing deep networks that combine multiple metamaterial elements and commonplace mechanical nonlinearities.

%Phononic resonators can attain quality factors of $10^9$, hence dissipating energies on the order of $10^{-27}$ joules per period of oscillation at common oscillation amplitudes [ref]. This stands in stark contrast with current electronic transistors that dissipate a minimum of $10^{-18}\;J$ [ref] to switch. Elastic systems can process signals directly in the mechanical domain domain, further improving energy efficiency by reducing the transduction loss.  Despite these unique characteristics, experimental demonstrations of elastic systems performing artificial-intelligence tasks are still lacking – with only a few theoretical results reported so far [Sylvestre, Fan]. The state of the art in phononic metamaterials is exemplified by topological models [cite Natures and Sciences], with a complexity around ten model parameters. In contrast, minimal machine learning models start at hundreds of parameters, reaching billions in the case of large language models. While significant advances have taken place in the combinatorial design of statically-deforming mechanical metamaterials, demonstrations of data-driven design of phononic metamaterials with sufficient complexity to perform artificial intelligence tasks is still missing.

\renewcommand{\topfraction}{1.9}
\renewcommand{\bottomfraction}{1.9}
\renewcommand{\textfraction}{.05}
\renewcommand{\floatpagefraction}{.9}
\renewcommand{\dbltopfraction}{0.95}
\renewcommand{\dblfloatpagefraction}{0.95}

%% file: 2_design.tex
\begin{figure*}
	%\centerin
	\includegraphics[width=0.95\linewidth]{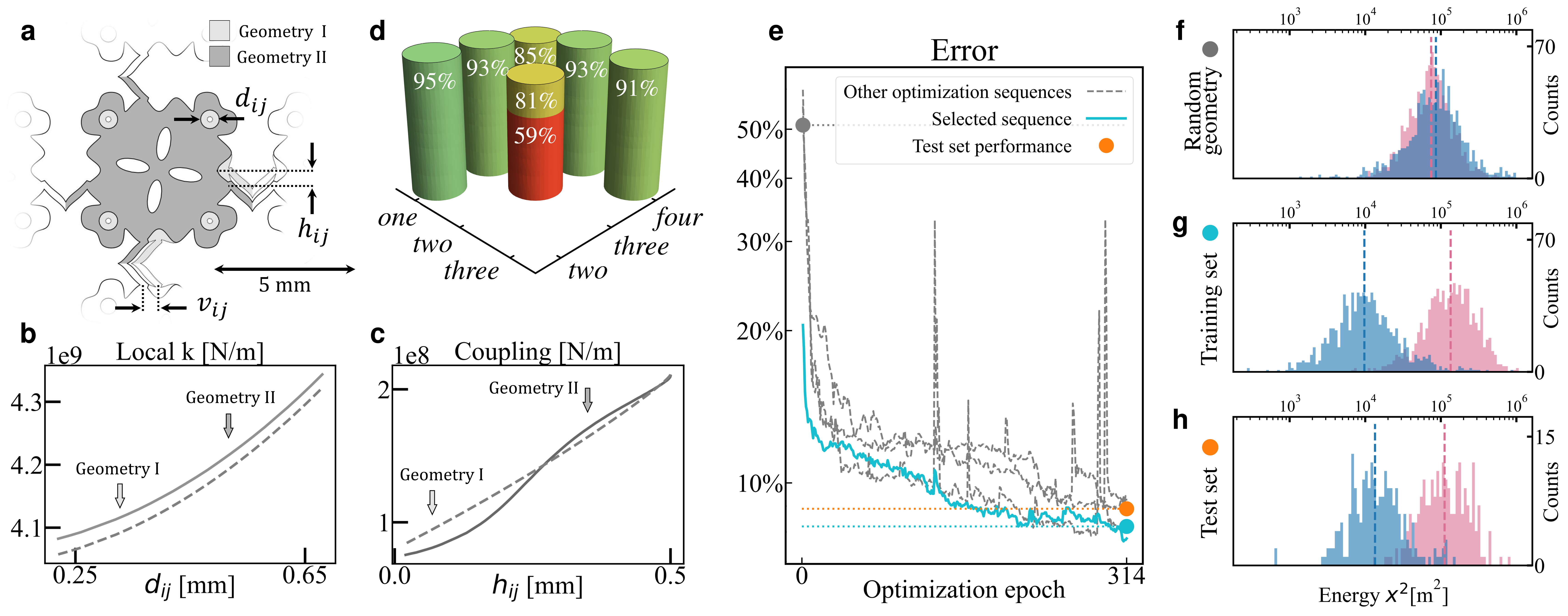}
\caption{ Sample design | 
{\fontfamily{phv}\selectfont\textbf{a}}
 Each unit cell $(i,j)$ contains four holes of equal diameter $d_{ij}$. The location of the coupling beams is parameterised by $h_{ij}$ and $v_{ij}$. We optimised the geometry by adjusting hole sizes and beam locations.
{\fontfamily{phv}\selectfont\textbf{b}}
Local stiffness and
{\fontfamily{phv}\selectfont\textbf{c}}~coupling strength as a function of the hole diameters $d_{ij}$ and beam locations $h_{ij}$, respectively. The approximation obtained by the surrogate model is shown with dashed lines. The coupling is strongly suppressed if the beam is attached where the plate eigenmode has a node, making a small beam displacement cause a large shift in the coupling constant~[Fig.~\ref{Fig1}c].
% A small beam displacement causes a large shift in the coupling constant, because the coupling is strongly suppressed when the beam is placed at the boundary centre, where the plate eigenmode has a node~[Fig.~\ref{Fig1}c]. 
The dark and pale grey arrows denote the points corresponding to the dark and pale shapes in a.
{\fontfamily{phv}\selectfont\textbf{d}}~ Speech classification accuracy for all pairs of spoken digits between {\it one} and {\it four}. For all but one of the pairs considered, a single layer provides a high classification accuracy. For the two-three word pair, performance can be increased from $59\%$ to $81\%$ by building a two-layer network (See generalisation section).
{\fontfamily{phv}\selectfont\textbf{e}}~Binary classification error rate evolution during training for  the {\it three-four} pair. The error rate is calculated on the training (lines) and test (dots) sets for the best-performing initial configuration. Training errors for other initial configurations are shown in gray. 
{\fontfamily{phv}\selectfont\textbf{f}}~Simulated binary classification performance of a structure with randomly chosen geometrical parameters $\left\{ d_{ij}, h_{ij}, v_{ij}\right\}$.
The optimized geometry is predicted to have an accuracy of
{\fontfamily{phv}\selectfont\textbf{f}}~$91.8\%$ on the training set and
{\fontfamily{phv}\selectfont\textbf{h}}~$91.1\%$ on the test set.
}
\label{Fig2}
\end{figure*}
%    \end{minipage}
%\end{widetext}
\bigskip

\textbf{Metamaterial design.} We considered a neural network  consisting of a phononic linear transformation implemented by a lattice of 7x7 sites [Fig.~\ref{Fig1}b], combined with a quadratic nonlinearity at the output site. For single-layer networks, the role of the lattice can be intuitively understood as a selective transmission of signal when excited by one word, but not another [Fig.~\ref{Fig1}c], while the nonlinearity can be seen as calculating the energy by squaring the displacement. State of the art phononic metamaterials, such as those realising topological insulators, are described by tight-binding models characterised by a few effective parameters ~\cite{Lin2022, Serra-Garcia2018, He2018, doi:10.1126/science.abj5488, delPino2022}. In contrast, machine learning models require hundreds to billions of parameters to encode a task. To bridge this expressibility gap, we combined a perturbative metamaterials approach~\cite{Matlack2018} with a surrogate model \cite{WHITE20191118} for gradient estimation: From the lattice geometry, we extracted an effective mass-spring model [Fig.~\ref{Fig1}c] through the Schrieffer-Wolff transformation (see \cite{PhysRev.149.491} and Supplementary Information). We parametrized the lattice so to obtain a high variability of the effective spring constants~[Fig.~2a-c] encoded by the fewest possible effective properties of the lattice geometry. Namely, only three quantities per lattice site were enough to yield a high variability of local resonant frequencies (hole diameters $d_{ij}$) and connection strengths (arm positions $v_{ij}$ and $h_{ij}$).

%\begin{widetext}[B!]
%    \begin{minipage}{\linewidth}

We trained the sample in-silico on utterances of spoken digits from the Google\textsuperscript{TM} Speech Commands Dataset~\cite{speechCommands}, using backpropagation in time with a sigmoidal loss function. The training objective was to produce a sample that accurately distinguishes between two chosen words, recorded from a large and diverse group of speakers under real-life conditions. The one-layer model consisting of a linear lattice with a quadratic readout function yielded classification performances above 90\% for the majority of tested word pairs [Fig.~\ref{Fig2}d]. The design process  started from a random configuration of the metamaterial lattice [Fig.~\ref{Fig2}f]. We then minimised the loss function using the Broyden-Fletcher-Goldfarb-Shanno (BFGS) algorithm~\cite{Battiti1990}, on batches containing the full training dataset. The gradients with respect to effective masses and springs were converted to gradients with respect to geometric parameters using the differentiable surrogate model.  The  optimisation process consisted of 300 iterations and was repeated for 15 different random initial designs. This process is shown in Fig.~\ref{Fig2}e for the \emph{three-four} word pair. Although full-batch BFGS has been associated with overfitting~\cite{fedorova2015exploring}, we observed an excellent generalisation performance\textemdash the degradation was less than 1\% between training and test datasets [Fig.~\ref{Fig2}g,h]. The optimised design that performed best on the training dataset for the \emph{three-four} word pair [Fig.\ref{Fig2}g] was selected for fabrication.

%% file: 3_realization.tex
\bigbreak
%\begin{widetext}
%    \begin{minipage}{\linewidth}
\begin{figure*}
	%\centering
	\includegraphics[width=0.95\linewidth]{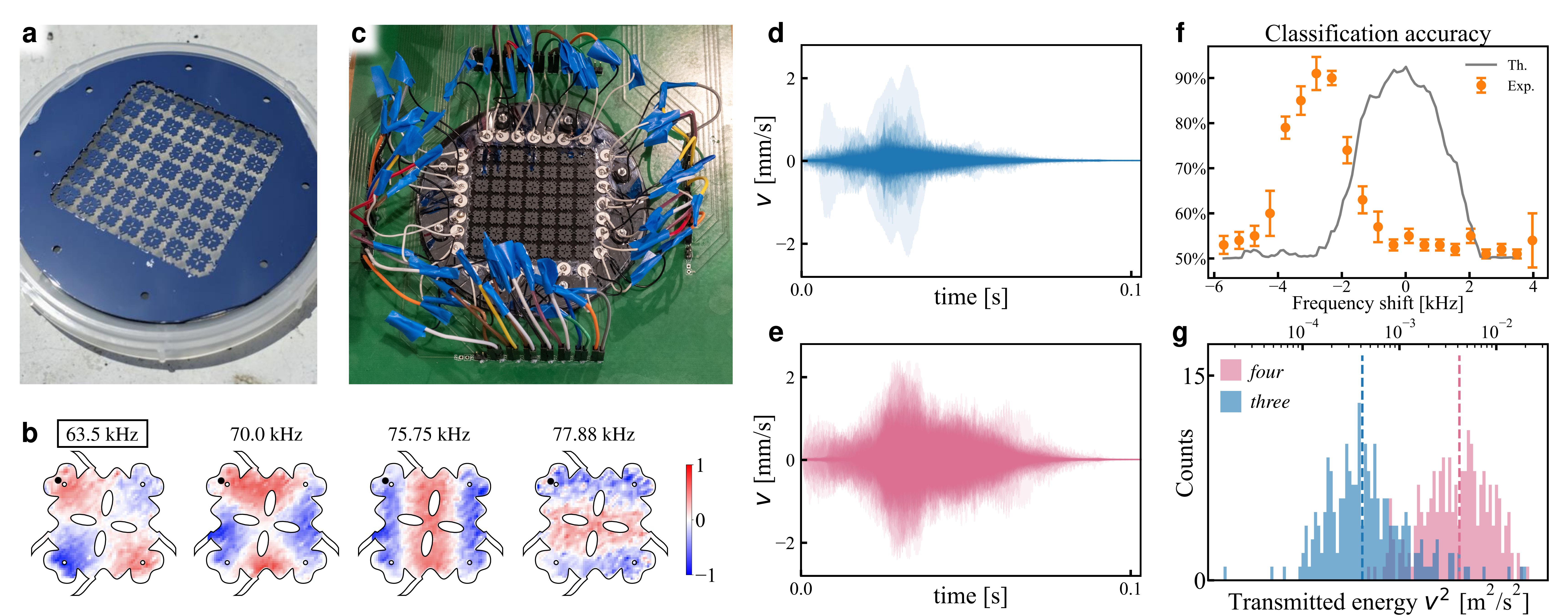}
\caption{ Experimental realisation | 
{\fontfamily{phv}\selectfont\textbf{a}}
Metamaterial lattice fabricated on a silicon wafer.
% {\fontfamily{phv}\selectfont\textbf{b}}
% Localized edge response measured at 56 points on the sample edge, after driving it by a single piezoelectric actuator. Later, all actuators are driven simultaneously with modulated recordings of spoken words, which provides a uniformly-moving sample boundary condition
{\fontfamily{phv}\selectfont\textbf{b}}
Measured plate vibrations under harmonic excitation at different frequencies. The black dot represents the point where the neural network output is taken.
{\fontfamily{phv}\selectfont\textbf{c}}
Experimental setup (photography by Astrid Robertsson).
{\fontfamily{phv}\selectfont\textbf{d}}
Measurements of the plate vibration at the output point (band-limited to $62.5-74.5$ kHz), superimposing the results for the excitation with each of the spoken \textit{three} and
{\fontfamily{phv}\selectfont\textbf{e}}
\textit{four} sound files in the training dataset. The signals corresponding to \textit{three} present a lower vibration amplitude. 
{\fontfamily{phv}\selectfont\textbf{f}}
Classification accuracy as a function of modulation frequency.
{\fontfamily{phv}\selectfont\textbf{g}}
Transmitted energy distribution, calculated from the individual curves in \textbf{d},\textbf{e}.
}
\label{Fig3}
\end{figure*}
%    \end{minipage}
%\end{widetext}

\textbf{Experimental realization}
We fabricated the sample~[Fig.~\ref{Fig3}a] on a $380\:\mu m$ silicon wafer using standard photolithography and etching techniques [Supplementary Information]. The equivalence between the full metamaterial and the mass-spring model, provided by the Schrieffer-Wolff transformation, depends on having an isolated phonon band. For materials without local potentials such as those compatible with our fabrication platform~\cite{Serra-Garcia2018}, this requires using a  high-order mode (Fig 3c), as the low-frequency spectrum is populated by the three degenerate bands arising from rigid translation modes. To map the broadband speech signal to a high-order band we modulated the speech on a $10.5 KHz$ carrier and then increased the playback speed by a factor of $6.8$. Such signal transformation would not be necessary for materials with a local support fabricated on multi-layer substrates [Supplementary Information], as the local support can be used to lift the degeneracy between rigid body modes and allows the metamaterial to directly operate on a speech signal.

To impose fixed boundary conditions, the wafer was clamped between two rigid frames and excited uniformly using 28 thickness-mode piezoelectric actuators~[Fig.~\ref{Fig3}b]. We measured the vibration of the output plate using a scanning Laser Doppler Vibrometer (LDV), band-limited over the range of $62.5-74.5$ kHz to minimise the influence of higher-order lattice modes~[Fig.~\ref{Fig3}c]. The measurements~[Fig.~\ref{Fig3}d,e] showed a significantly larger center plate vibration when the lattice was excited by a \textit{four}\textemdash even though all excitation signals were normalised to the same mean energy. The optimal classification accuracy was obtained when the modulation frequency was shifted by $2.8$  kHz~[Fig.~\ref{Fig3}f] with respect to the design value. This deviation can be accounted by the manufacturing tolerance in the thickness of the wafer, which is nominally
$\pm 10\:\mathrm{\mu m}$, and can be corrected by combining the theoretical model with physical measurements~\cite{Wright2022} to trim the sample after fabrication~\cite{Hsu2007FrequencyTF}. With the optimal modulation frequency as determined on the training set~[Fig.~\ref{Fig3}f], we measured a test-set classification accuracy of $89.6\%$~[Fig.~\ref{Fig3}g], close to the simulated value of $91.1\%$.

%% file: 4_outlook.tex
% \section{Outlook}

\begin{figure*}
	%\centering
	\includegraphics[width=0.95\linewidth]{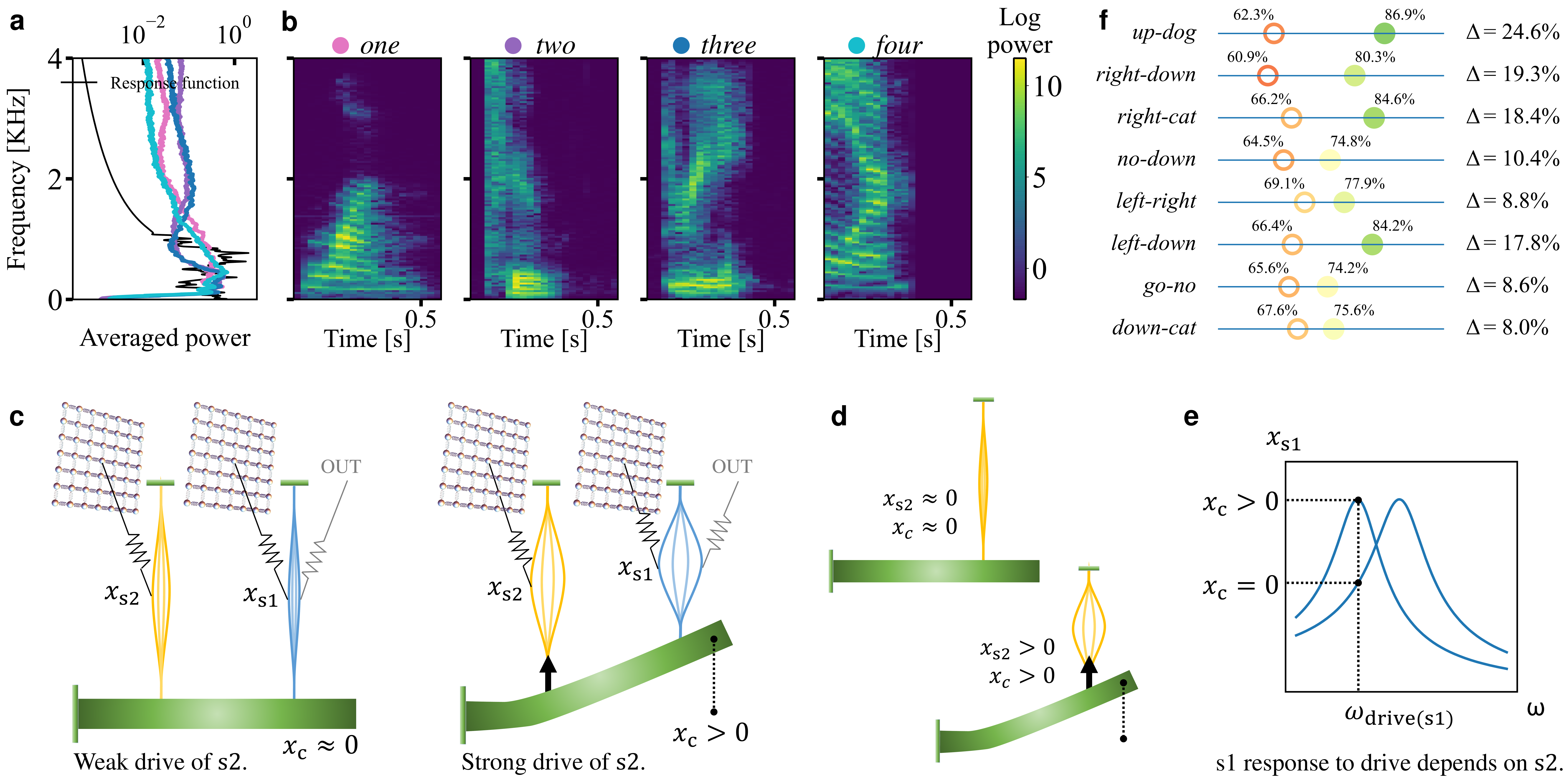}
\caption{Generalisation to other word pairs |
% {\fontfamily{phv}\selectfont\textbf{a}}
% Theoretical results show that high-performance linear metamaterials can be designed for all considered pairs of numbers, except for \textit{two-three}. Performance can be further improved by combining linear metamaterials ('weights') with fixed nonlinear elements ('activation functions'). The improvement as a consequence of adding a single nonlinear term is shown for the {\it two-three} pair.
{\fontfamily{phv}\selectfont\textbf{a}}
Mean frequency content of the words {\it one}, {\it two}, {\it three} and {\it four} (pink, purple, blue and magenta, respectively), and transfer function of the linear lattice designed to distinguish between {\it three} and {\it four}. Word pairs with more distinct mean frequency contents can be classified more accurately by a single-layer device.
{\fontfamily{phv}\selectfont\textbf{b}} Example spectrograms for the words {\it one}, {\it two}, {\it three} and {\it four}. Word pairs with similar frequency contents can be distinguished from the temporal ordering of the frequency components. This distinction can be mechanically implemented through multi-layer (deep) networks.
{\fontfamily{phv}\selectfont\textbf{c}}
Two layer network implemented by combining two linear transformations interacting through a {\it mechanical nonlinear activation function}, consisting of two strings ($\mathrm{s1}$, $\mathrm{s2}$) and a cantilever ($\mathrm{c}$), thus realising an asymmetric quadratic nonlinearity.
{\fontfamily{phv}\selectfont\textbf{d}}
When $\mathrm{s2}$ vibrates with high amplitude it is on average more curved, and hence deflects the cantilever $\mathrm{c}$ due to its finite stretching compliance (force denoted by thick black arrow). 
{\fontfamily{phv}\selectfont\textbf{e}}
The time-dependent position of the cantilever $\mathrm{c}$ then influences the tension of the string $\mathrm{s1}$, shifting its resonance curve and altering the final output $x_\mathrm{s1}$.
{\fontfamily{phv}\selectfont\textbf{f}}
The string-cantilever-based nonlinearity significantly improves the classification accuracies for all tested word pairs with similar spectral content.
}
\label{Fig4}
\end{figure*}

\textbf{Interpretation and generalisation} The full phononic metamaterial is interpreted as a single linear transformation that, when coupled with a nonlinear activation function, implements a layer of a neural network. The action of the metamaterial on the input signal can be understood as a convolution between the speech signal and a kernel encoded in the impulse response of the lattice.  Although the lattice contains only nearest-neighbor interactions, the linear transformation effected by the lattice is dense in time, with the weights for long-range temporal interactions determined by integrating all possible paths that sound waves can take through the lattice with a given signal delay. The effect of the training process is to optimise the weights associated to each delay. Convolution by an impulse response kernel is equivalent to applying a frequency filter with the transfer function, the Fourier transform of the impulse response. This provides a direct interpretation to the classification capabilities of the single lattice: During the design process the lattice learns to maximize its energy transfer at the frequencies where the difference between words is maximal (Fig.~\ref{Fig4}a). The quadratic nonlinearity then rectifies this selectively-transferred signal and computes the mean energy. This mechanism allows the passive metamaterial to distinguish between linearly separable word pairs.

%\begin{widetext}
%    \begin{minipage}{\linewidth}

%    \end{minipage} 
%\end{widetext}

Passive mechanical speech classification can be generalised to word pairs with similar mean spectral contents by assembling deep networks interconnected by nonlinear elements [Fig. 1b]. These nonlinear elements allow the lattice to distinguish the temporal ordering of different frequency components. We optimise a deep network consisting of two 7x7 mass-spring lattices interconnected with the nonlinear mechanical element from Reference \cite{Serra-Garcia2016}. This nonlinear element consists of two strings connected to a cantilever. Due to geometric effects, the vibration of strings results in a dynamic increase of their tension, which deflects the cantilever [Fig~\ref{Fig4}c]. In turn, the deflection of the cantilever dynamically alters the tension of the strings, shifting their resonance frequency [Fig.~\ref{Fig4}d]. This nonlinear mechanism can be interpreted analogously to a gating mechanism in conventional recurrent speech models \cite{chung2014empirical}; the cantilever-induced shift in frequency of the string modulates the flow of information between lattice and output by altering the alignment between their respective resonance frequencies. A two-layer model more than halved the classification error, from $41\%$ to $19\%$, for the word pairs {\it two}-{\it three}. Significant improvements were obtained in all tested word pairs with similar spectral content [Fig.~\ref{Fig4}f]. 

The theorem by Boyd and Chua \cite{boydchua1985fading}, guarantees that mechanical systems can theoretically reach accuracies comparable to those of electronic systems, as any fading-memory function can be realised as a combination of linear transfer functions and static nonlinearities. Speech recognition is by definition fading memory -- the result cannot depend on signals that took place before the duration of the detected word; arbitrary linear transfer functions can be engineered by branched delay lines; and arbitrary static nonlinearities can be realised by cascading quadratic elements.

 By demonstrating that machine learning tasks can be encoded in the response of phononic metamaterials, together with prior work on passive amplitude activated switches~\cite{AmpliSwitch}, we illuminate a novel path towards zero-power smart devices that can intelligently respond to events. This capability is out of reach of conventional electronics: State-of-the-art transistors require more than $10^{-18}$ Joules to switch~\cite{doi:10.1126/science.ade7656}. In contrast, phononic resonators can easily go below $10^{-21}$ Joules per period of oscillation~\cite{Sazonova2004}. This potential for orders-of-magnitude improvement in energy efficiency had already been recognised in the context of conventional digital computing~\cite{doi:10.1126/science.1144793} and can now be applied to machine learning problems.

%% file: 5_appendix.tex
\cleardoublepage

\section{Finite element method for structure simulation}
\label{SI:FEM}

The analysis of the dynamics of the phononic system starts with the finite element method (FEM) simulations. FEM is a commonly used method for numerically solving partial differential equations, such as the three-dimensional linear elasticity equation. The continuous elastic medium (the lattice of resonators) is discretised in a set of finite elements, so as to enable a numerical simulation of its response. The equations of motion for the nodes in these finite elements and the corresponding boundary conditions are collected in a set of equations that can be written in a matrix form,
\begin{equation}
	M { \ddot{\vec{x}}} + K \vec{x} = {\vec{f}}(t) ,
\end{equation}
where ${\vec{x}}$ are the positions of the nodes in the finite elements, $\vec{f}(t)$ is the force acting on them, and the matrices $M$ and $K$ reflect their mass and stiffness properties.
The size of the mass and stiffness matrices, and of the system of equations that needs to be solved in order to extract the system dynamics, directly depends on the discretization density. Relatively high precision is required in order to succeed in the speech recognition task, on the order of $10^7$ degrees of freedom for the $7\times 7$ lattice. This motivates the approach via dynamic sub-structuring, as explained in the next section.

%%%%%%%%%%%%%%%%%%%%%%%%%%%%%%%%%%%%%%%%%%%%

\section{Dynamic sub-structuring}
\label{SI:substructuring}

Complex signal processing tasks such as speech recognition require tightly-controlled mechanical properties. Therefore, it is necessary to perform FEM simulations using a very fine mesh. In this work, we use 323 kDOFs per site, which adds up to roughly 16 million degrees of freedom for the total design. Such a large number of degrees of freedom presents a computational challenge, especially given that we must perform many full lattice simulations to complete a device optimization.

\begin{figure}[H]
\centering
\includegraphics{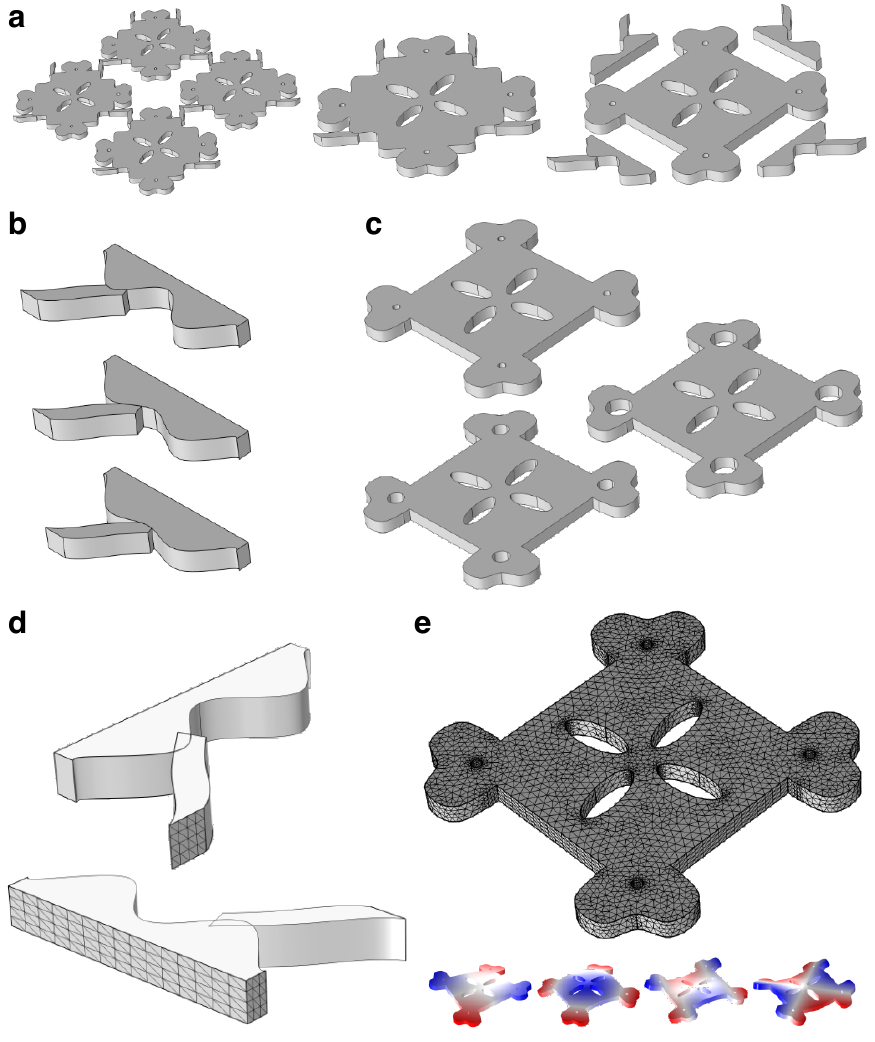}
\caption{Simulating a lattice by dynamic substructuring | \textbf{a} The lattice is divided in sites, which are then divided into link and resonator components. \textbf{b, c} The links (b) and resonators (c) are chosen from a library of pre-computed building blocks.  \textbf{d} The full model is assembled by imposing matching displacements at the boundary meshes. \textbf{e} The internal (non-boundary) degrees of freedom [top] are replaced by a set of eigenmodes [bottom]. Note that not all retained eigenmodes are shown here. }
\label{fig:substructure}
\end{figure}

We address the problem of having a large number of DOFs by performing dynamic sub-structuring \cite{CraigBampton}. A lattice consisting of multiple vibrating plates is assembled from individual plates, which in turn are assembled from building blocks such as links and hole designs~[Fig.~\ref{fig:substructure}a]. While the number of plate designs is very large ($>2.6\cdot10^9$), the number of different building blocks is small: $100$ instances for the arms and $26$ instances for the plate hole pattern. This allows us to pre-compute lumped models for those plate designs~[Fig.~\ref{fig:substructure}b,c]. Besides the time savings obtained by bypassing the meshing and matrix assembly steps at every iteration, using a pre-computed dynamical matrices allows for increased computational parallelisation. While FEM simulation is limited by the number of available software licenses, the combination of pre-computed building blocks can be performed using open-source tools and therefore does not pose an obstacle to paralellisation.

To perform dynamic sub-structuring, the finite element models for every component are divided into bulk and boundary degrees of freedom. The full model is assembled by imposing matching displacements between the degrees of freedom at the interfaces between components [Fig.~\ref{fig:substructure}d]. The bulk degrees of freedom of each building block are replaced by lattice eigenmodes obtained under free boundary conditions, following the approach by Rubin ~\cite{RubinMethod}. We observed that the Rubin approach, with eigenfrequency errors smaller than $0.1$ \% significantly outperformed the Craig-Bampton method~\cite{CraigBampton}, in which the bulk DOFs are replaced by eigenmodes of a structure under fixed conditions at the interface degrees of freedom, resulting in an eigenfrequency error above $1$ \%. This is consistent with prior reports on the low-frequency modelling of structures~\cite{DomainDecomposition}. 

In this work, we model the bulk DOFs by retaining $75$ eigenmodes per resonator site. The interfaces are modelled using $972$ DOFs. Since interface DOFs are shared between neighboring sites, each plate requires $561$ individual DOFs, down from the $323$ kDOFs required when performing direct FEM simulation.  It should be noted that the dynamic matrices obtained by dynamic substructuring are much less sparse than those from direct finite element modelling, limiting the observed speedup of a full system diagonalisation to a factor of approximately 10.
To ensure matching meshes at the interfaces between different components, we used COMSOL Multiphysics\textsuperscript{\textregistered} explicit mesh functions. In complex geometries combining parametric curves and boolean operations, we observed fluctuations in the location of the boundary mesh DOFs even when the explicit mesh function was used. Using a boundary layer mesh function to improve resolution around the interface areas mitigated the problem.

\section{Effective mass-spring model extraction}
Finite element models contain a very large number of DOFs. However, the dynamics of the system at the range of frequencies of interest can be described by the small set of normal modes whose resonance frequencies lie in the relevant frequency range. For the lattices and frequency ranges considered in this work, a modal description requires one DOF per site, or $49$ DOFs for the entire lattice.

\begin{figure}[H]
\centering
\includegraphics{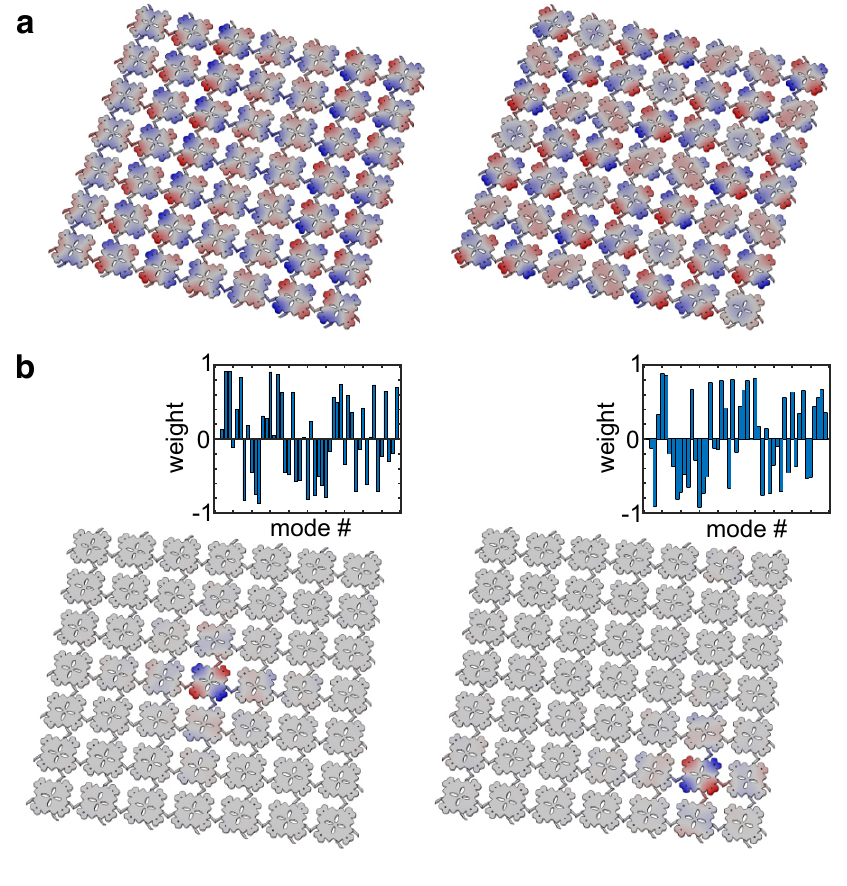}
\caption{Assembling a localised basis from lattice modes | \textbf{a} Example showing two lattice eigenmodes in the frequency range of interest. Lattice eigenmodes are delocalised. \textbf{b} By combining the 49 delocalised modes with suitable weights (histogram), we can assemble highly localised basis elements.}
\label{fig:locmodes}
\end{figure}

While normal modes [Fig.~\ref{fig:locmodes}a] provide a very efficient description to simulate the dynamics of the lattice, they form an unsuitable basis for optimization. This is because the properties governing them (effective mass, stiffness and coupling force) depend on the full lattice geometry. Such full-lattice dependence has two negative consequences: First, it is very difficult to build a surrogate model that can be used to compute gradients or rapidly estimate the performance of a particular geometry. Second, determining the normal modes requires a full-lattice diagonalisation, which scales quadratically with the number of DOFs, and therefore limits the sizes of systems that can be designed. Furthermore, since normal modes extend through the whole lattice, the memory requirements for storing a normal mode basis also increase quadratically with system size, as both the number of modes, and size of the modal functions increase with lattice size.

These two disadvantages of normal modes can be avoided by describing the dynamics of the system in terms of DOFs localized at every site [Fig.~\ref{fig:locmodes}b], analogous to atomic orbitals in Density Functional Theory. In this representation, each element of the dynamic representation depends mostly on the geometric parameters close to the site where the corresponding DOF is localized. Since the geometry surrounding any particular site can be described by a small number of parameters, the resulting simple dependence can be readily captured by a surrogate model that we use to estimate gradients and as replacement of full lattice simulations. Furthermore, since localized basis functions decay very rapidly, truncating them at a fixed radius allows for linear memory storage of basis functions.

Throughout this work, we compute the maximally localized basis by calculating the normal modes of the full lattice $\psi_i$, selecting those modes that lie in the frequency range of interest, and then finding the linear combinations of those modes that are maximally localized at every site $j$:

$$\phi_j = \sum_{i}a_{ji}\psi_i$$

Where the coefficients $a_{ji}$ are found by solving the maximization problem 
$$\max_{a_{ji}}\left[ \phi_j^T P_j \phi_j\right],$$
where $P_j$ is a projector with value $1$ for the degrees of freedom in site $j$ and $0$ for all other degrees of freedom. The maximization is performed under the normalization condition

$$\sum_i{a_{ji}^2}=1.$$

Once the maximally-localised basis elements $\phi_i$ have been determined, the mass $M$ and stiffness matrices $K$ for the reduced mass-spring model can be determined from the FEM mass and stiffness matrices by performing the rotation $K = \Gamma ^T K_{\textrm{FEM}} \Gamma$ and $M = \Gamma ^T M_{\textrm{FEM}} \Gamma$ where $\Gamma$ is a matrix whose columns are the basis elements $\phi$, $\Gamma={\phi_1, \phi_2, ..., \phi_n}$.

The change of coordinates introduced by $\Gamma$ has the effect of isolating the dynamics of the system that takes place at the frequency range of interest, and plays an analogous role to the transformation $e^S$ introduced by Schrieffer and Wolff~\cite{PhysRev.149.491} that we had used in prior works \cite{Matlack2018}:

\begin{equation}
e^{S}K_{\textrm{FEM}}e^{-S}=
\begin{pmatrix} 
K & 0 \\ 
0 & K_I
\end{pmatrix} ,
\end{equation}

\noindent where $S$ is the generator of the transformation \textemdash computed as a series expansion \cite{BRAVYI20112793}, and $K_I$ represents a stiffness matrix aggregating all irrelevant degrees of freedom (i.e., whose dynamics takes place outside the range of frequencies of interest). This analogy allows an alternative method to compute the effective theory based on a series expansion \cite{BRAVYI20112793, Matlack2018}, where the interaction between lattice sites is considered a small parameter. The key advantage of this approach is that it leads naturally to linear-time algorithms to determine the effective theory~\textemdash as every site dynamics depends only on a fixed number of neighbors. However, the approach  presents significant challenges due to the poor convergence of the series expansion~\cite{VariationalSW}.

%%%%%%%%%%%%%%%%%%%%%%%%%%%%%%%%%%%%%%%%%%%%

\section{Mode selection and site design}
\label{SI:modeSelection}

\begin{figure}[t]
\centering
\includegraphics{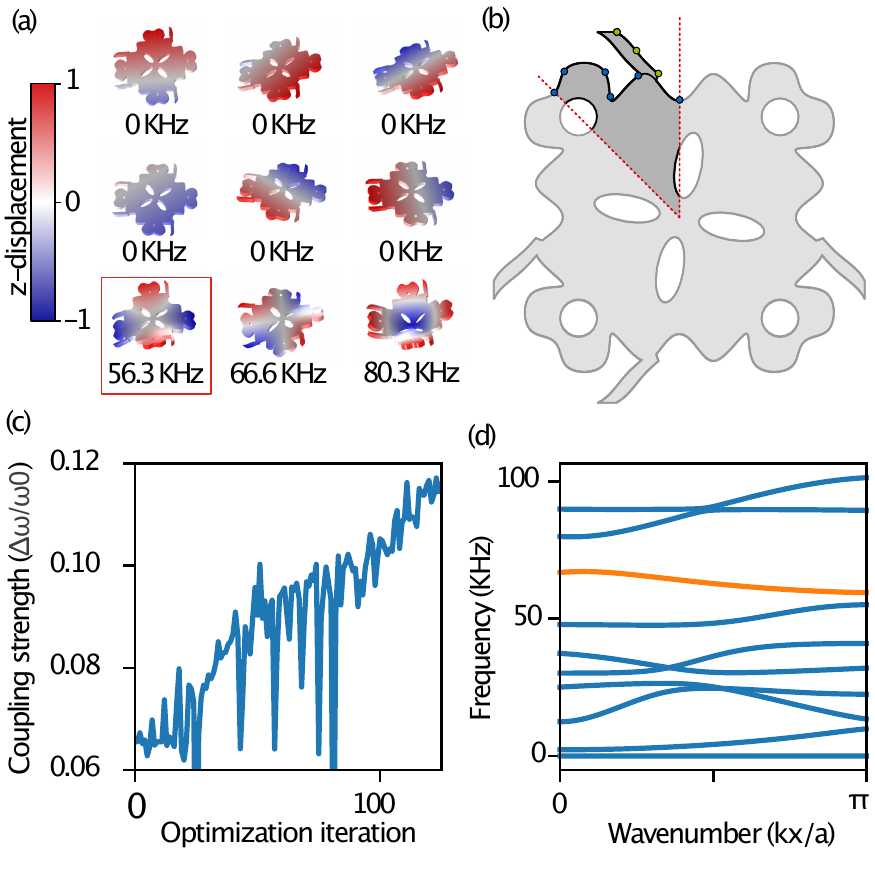}
\caption{Optimisation of the site design | \textbf{a} First 9 normal modes of a free-standing site. The red rectangle highlights the selected mode. \textbf{b} The site design is parameterized by a set of control points (blue dots) together with the radial dimension, rotation angle and radial location of the center elliptic holes. The coupling beams are parameterized by control points (green dots) together with a thickness parameter. \textbf{c} Evolution of the coupling strength during the site optimisation. \textbf{d} Band structure of a periodic material consisting of an optimised unit cell. The orange band arises from the selected normal mode.}
\label{fig:combi_celloptimisation}
\end{figure}

To be able to efficiently model the lattice as a mass-spring model, the design must satisfy certain conditions. First, it is necessary that every lattice site has only one vibrational normal mode in the frequency band of interest. Second, it must be possible to implement inter-site couplings of different strength values with realistic geometric parameters. Under the perturbative (weak inter-plate coupling) assumption, the normal modes of the lattice can be traced to the modes of the freely-vibrating plates~\cite{Matlack2018}~[Fig.~\ref{fig:combi_celloptimisation}a]. The first non-rigid-body mode is a good candidate because its boundary vibration profile alternates between positive and negative displacements and has a node. Therefore, by placing the linking beams closer or farther from the node, it is possible to reproduce a broad range of couplings.

To maximise the design space of effective mass-spring model parameters, the plates and coupling beams must be designed to allow for large inter-plate coupling strengths, while maintaining a good separation between various modes of resonance. Such separation is critical as it determines the maximum bandwidth that the structure can handle, while still being represented accurately by a simple mass-spring model~\cite{Matlack2018}. These properties are attained by numerical optimisation. We define the contour of the plates and coupling beams using splines parameterised by a set of control points~[Fig.~\ref{fig:combi_celloptimisation}b]. Then, we perform an optimisation that maximises the interaction strength~[Fig.~\ref{fig:combi_celloptimisation}c]. The optimisation is interrupted early to prevent the design from reaching extreme geometries (e.g. very thin beams) that would be hard to fabricate.
The plate optimisation is performed using the Nelder-Mead algorithm, because it does not require gradients of the misfit, which would be hard to obtain for a spline parameterisation. To compute the misfit, we consider a periodic geometry where the site is subject to Bloch boundary conditions with wavenumber $k$. For this configuration, we can define a band structure~[Fig.~\ref{fig:combi_celloptimisation}d] -- it should be noted that in the periodic configuration, the seventh free-plate mode actually manifests as the eighth band.

The misfit for optimisation is determined by evaluating the band structure at four equally-spaced points in the Brillouin zone. Then, the spectral separation is defined as $$G=\sum_k{\frac{1}{f_8(k)-f_7(k)} + \frac{1}{f_9(k)-f_8(k)},}$$ where $f_i$ is the frequency of the $i$'th mode at wavenumber $k$. The bandwidth is defined as $$\Delta \omega = \sum_k{f_8(k)\cos (k)},$$ and the center frequency is defined as  $$\omega_0 = \frac{1}{4}\sum_k{f_8}(k).$$ With these parameters, we observe that minimising a misfit of the form $$N = -800\frac{\Delta \omega}{\omega_0} + G + \frac{1}{64}G^2$$ produces designs with a good balance between coupling strength and frequency separation.

%%%%%%%%%%%%%%%%%%%%%%%%%%%%%%%%%%%%%%%%%%%%

\section{Time-reversal gradient calculation}

The evolution of a mass-spring system can be described by an equation of the form$$\frac{d\vec{r}(t)}{dt}=\vec{f}\left(\vec{r}(t), \vec{p}, t\right),$$ where $\vec{r}=(x_1, x_2, ..., x_n, v_1, v_2, ..., v_n)$ is a state vector containing the positions $x_i$ and velocities $v_i$ of the masses in the system, $\vec{p}$ is a vector with the parameters of the system (masses, springs, damping factors), and $t$ is the time. 

For every audio sample, the energy transmitted is a functional of the state evolution $\vec{r}(t)$. Most neural network optimisation algorithms require access to the gradient of the output to efficiently identify the system parameters that lead to the desired response. A naive approximation to calculate this gradient involves forward-integrating the Jacobian of the state vector $\vec{r}$ with respect to the system parameters, $J_p$. However, such an approach involves solving as many Ordinary Differential Equations (ODEs) as the number of parameters that the system has. An alternative approach consists of determining an adjoint field $\vec{u}(t)$, which quantifies the dependence of the energy functional on a small perturbation of the state vector $d\vec{r}$, introduced at time $t$~\cite{automatedDesignDynamicalDambre}. The adjoint field $\vec{u}(t)$ can be calculated by solving a single additional ODE, $$\frac{d\vec{u}}{dt}= J_{\vec{r}}^T\left(\vec{r}(T-t), \vec{p}, T-t\right)\vec{u}+\nabla_{\vec{r}(T-t)}{E}.$$ The gradient of $E$ with respect to every parameter $p_i$ is given by $$\int_0^T{\vec{u}\cdot\frac{d\vec{f}}{dp_i}dt}$$.

Calculating the gradient using adjoint methods requires storing all dynamical variables $\vec{r}(t)$ at every time step. This is very memory intensive, and for this reason we divide the total number of time steps $N$ into $\sqrt{N}$ intervals. We perform one forward simulation, storing the dynamical variables at the start of every interval. During backward integration, every time the simulation goes through an interval boundary, $\vec{r}$ is calculated for all the $\sqrt{N}$ timesteps in the new interval. This has the effect of dividing the memory footprint by a factor of $\sqrt{N}$ while increasing the computation time by one forward simulation. In practice, the time increase is lower due to the reduction in memory bandwidth, as the buffers now fit in the CPU caches.

Both forward and backward simulations are performed using a fourth-order Runge-Kutta algorithm~\cite{press1988numerical} with a time step of $624.7$ ns.

%%%%%%%%%%%%%%%%%%%%%%%%%%%%%%%%%%%%%%%%%%%%

\section{Surrogate model for the lattice effective theory}
\label{SI:machineLearning}

\begin{figure}[b]
\centering
\includegraphics[width=0.5\columnwidth]{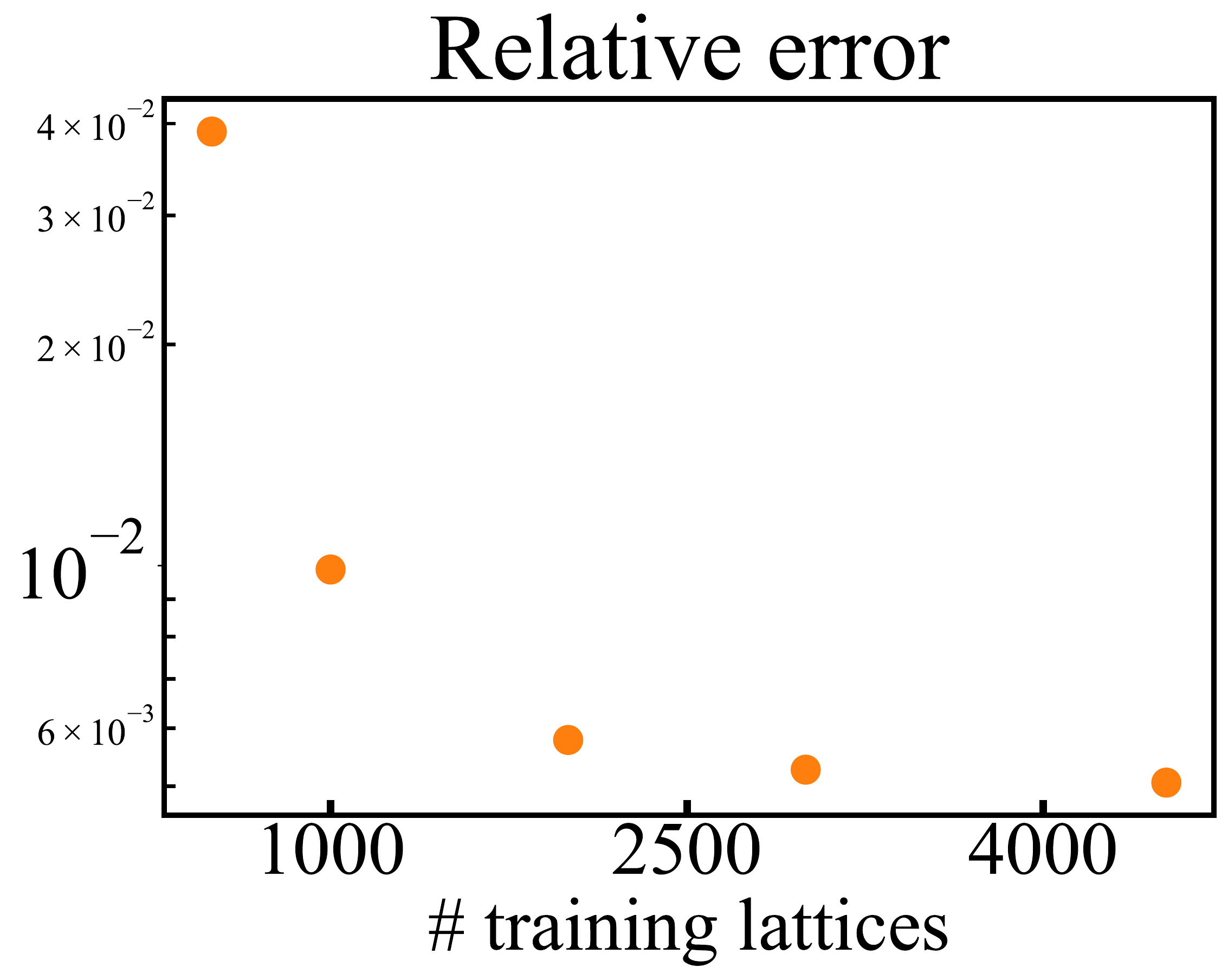}
\caption{Training of the surrogate model | Frobenius norm of the error between the estimated and simulated matrices $\left\lVert K_{\mathrm{FEM}}-K_{\mathrm{ML}}\right\rVert$, for a test set of 45 random geometries. }
\label{fig:mlmodel}
\end{figure}
We found it valuable to approximate the mass and stiffness matrices separately, instead of directly predicting the dynamic matrix $D=M^{-1} K$. This preserves the symmetry of $M$ and $K$ and ensures that the eigenvalues of $M_\mathrm{ML}^{-1} K_\mathrm{ML}$ remain real. In contrast, direct learning of the dynamic matrix resulted in complex eigenvalues.

Evaluating the mass-spring dynamic matrices using FEM simulations is very expensive, as we must assemble and diagonalise matrices involving millions of degrees of freedom. To solve this problem, we construct a surrogate model to approximate the system response. The surrogate model predicts the matrices as a function of the geometric parameters.

We express every element of the stiffness matrix K as $$K_{ij}=a_{ij}^K+\vec{b}^K\cdot\vec{p}_r + \vec{p}_r^T C_{ij}^K\vec{p}_r.$$ The equation for the mass matrix has the same form, substituting $K$ by $M$. Here, $a^{\{\mathrm{M},\mathrm{K}\}}_{ij}$, $\vec{b}^{\{\mathrm{M},\mathrm{K}\}}_{ij}$ and $C^{\{\mathrm{M},\mathrm{K}\}}_{ij}$ are scalar, vector and matrix parameters respectively, and are different for every matrix element ${\{M,K\}}_{ij}$. The vector $\vec{p}_r$ contains the subset of geometric parameters (hole radius and beam locations) that is most relevant to the particular matrix element ${\{M,K\}}_{ij}$. 

The relevant parameters associated to a matrix element $\{M,K\}_{ij}$ are identified by geometric proximity to the degrees of freedom $i$ and $j$. Given that degrees of freedom correspond to vibrations localised at an individual plate, we can assign a distance between every geometric feature and degree of freedom. For a degree of freedom $i$, holes in the plate corresponding to that degree of freedom, and locations of beams that are attached to this plate, are labelled as distance $0$, holes in a nearest neighbor, and beams incident to it are labelled as distance $1$, while holes and beams attached to a next-nearest neighbor are assigned a $2$. Elements that couple multiple sites (e.g. beams) are assigned the smallest of the two possible values. For a matrix element $\{M,K\}_{ij}$, geometric parameters corresponding to features that are at a distance of two or less are kept as relevant.

The system is trained on 5000 randomly generated training lattices, with values obtained from full FEM simulation.  We tested its accuracy  on a test set of 45 additional lattices, showing small residuals between predicted and simulated values~[Fig.~\ref{fig:mlmodel}d].
Using a surrogate model instead of a full FEM simulation drastically improves the optimisation speed. However, the resulting M and K matrices have limited accuracy. During optimisation, we address this problem by performing a full-lattice FEM simulation every 30 BFGS iterations. We the shift the predictions of the surrogate model by $\Delta K = \alpha (K_{FEM} - K_{Surrogate})$ and $\Delta M = \alpha (M_{FEM} - M_{Surrogate})$. The parameter $\alpha$ is introduced to prevent oscillations; as the optimisation progresses, $\alpha$ is increased smoothly from 0 to 1. The final training and test-set simulations are computed on a mass-spring model extracted from a full FEM simulation. The combination of surrogate model predictions with infrequent but exact FEM simulations results in a highly accurate final design.

%%%%%%%%%%%%%%%%%%%%%%%%%%%%%%%%%%%%%%%%%%%%
%

% \section{Extension to other word pairs}
% \label{SI:OtherWords}

% The highly performing binary classification by the passive elastic system is by no means bound to the chosen four digits, and can be extended to other word pairs. We show it by considering all $45$ pairs of ten common English words:\\

% \indent {\it bird, cat, dog, down, go, left, no, right, up, yes}.\\

% \noindent We optimize a generic mass-spring system, consisting of $7\times 7$ masses and connecting springs, to perform an efficient speech classification task for each of these $45$ word pairs. We train the system by backpropagation, starting from multiple random configurations of the parameters. The performances for the best performing configurations of the linear mass-spring lattice are shown in Fig.~\ref{fig:manyWords}. As expected, a significant fraction of word pairs can be linearly classified with accuracies around $90\%$, and for all but eight word pairs we achieve $>70\%$ binary classification accuracies. The improvement of the classification performance with nonlinear elements is discussed in Fig.~\ref{Fig4}a,d of the main text, and in the next two sections of the supplementary materials.

% \begin{figure}[b!]
% 	\centering
% 	\includegraphics[width=0.85\columnwidth]{Figures/ManywordsLinear.pdf}
% \caption{Generalisation to other word pairs |  Test classification accuracies (in percentage) of a linear $7\times 7$ mass-spring system for all pairs of ten common words.
% }
% \label{fig:manyWords}
% \end{figure}

%%%%%%%%%%%%%%%%%%%%%%%%%%%%%%%%%%%%%%%%%%%%

\section{Mechanical nonlinearities and deep networks}
\label{SI:nonlinear}

\begin{figure}[b!]
	\centering
	\includegraphics[width=0.95\columnwidth]{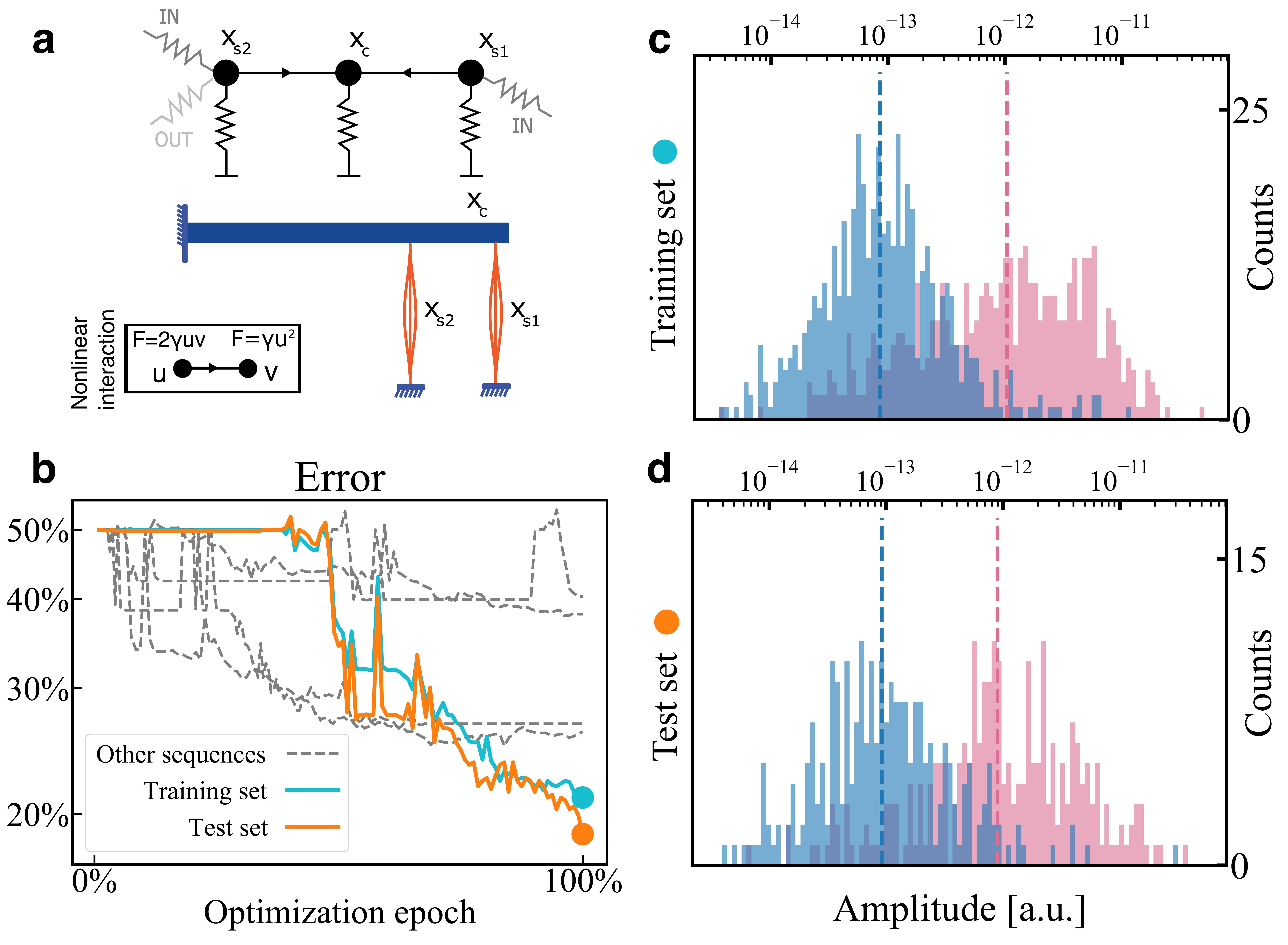}
\caption{Speech classification including a nonlinear element |  
{\fontfamily{phv}\selectfont\textbf{a}}
We use a system of two vibrating ribbons and a cantilever to nonlinearly combine the output from two lattices.
{\fontfamily{phv}\selectfont\textbf{b}}
Evolution of the classification accuracy during training, for different random initial values of the parameters.
{\fontfamily{phv}\selectfont\textbf{c,d}}
Histogram of the transmitted energy for the optimized classifier between \textit{two-three} for the training and test sets respectively.
{\fontfamily{phv}\selectfont\textbf{e}}
Performance of the system with a single nonlinear term (cyan in b) on the test set. Overfitting is expected due to the relatively small number of training samples as compared to the number of parameters in the nonlinear models.
}
\label{fig:nonlinearRealistic}
\end{figure}

In figure 4 of the manuscript, we show how deep networks consisting of linear lattices interacting through nonlinear mechanical elements~\cite{Serra-Garcia2016} can passively classify word pairs that are not linearly separable. The nonlinear element  considered [Fig.~\ref{fig:nonlinearRealistic}a], consists of two string-like resonators ($\mathrm{s1}$ and $\mathrm{s2}$) and a cantilever ($\mathrm{c}$), is described by the following equations of motion:

\begin{gather}
m_{s1} \ddot{x}_\mathrm{s1} + b_\mathrm{s1} \dot{x}_\mathrm{s1} + \left( k_\mathrm{s1} + 2\gamma_\mathrm{s1|c} x_\mathrm{c} \right) x_\mathrm{s1} = f_\mathrm{s1} \\
m_\mathrm{s2} \ddot{x}_\mathrm{s2} + b_\mathrm{s2} \dot{x}_\mathrm{s2} + \left( k_\mathrm{s2} + 2\gamma_{s2|c} x_\mathrm{c} \right) x_\mathrm{s2} = f_\mathrm{s2} \\
m_c \ddot{x}_\mathrm{c} + b_\mathrm{c} \dot{x}_\mathrm{c} + \left(k_\mathrm{c} + \gamma_\mathrm{s1|c} x_\mathrm{s1}^2 + \gamma_\mathrm{s2|c} x_\mathrm{s2}^2\right) x_\mathrm{c} = f_\mathrm{c} ,
\end{gather}

with effective masses $m_\alpha$, damping $b_\alpha$ and linear stiffnesses $k_\alpha$, where $\alpha=\{\mathrm{s1},\mathrm{s2},\mathrm{c}\}$. The nonlinear terms describe the interaction between the string tension and the cantilever displacement: When the cantilever moves, it increases the tension of the strings causing their frequency to shift upwards. At the same time, the length of the strings increases with its flexural deformation (deformed strings are longer due to curvature). This effect manifests as a force on the cantilever proportional to the square displacement of the strings. The nonlinear constants $\gamma_{s1|c}$ and $\gamma_{s2|c}$ govern the nonlinear interaction between strings and cantilever.

We train the system of [Fig.~\ref{Fig4}c]) using gradient descent [Fig.~\ref{fig:nonlinearRealistic}b], reaching an accuracy of $78\%$ and $81\%$ for the training and test set respectively [Fig.~\ref{fig:nonlinearRealistic}c,d]. The slightly increased performance of the test set can be attributed to statistical fluctuations due to the relatively small sample size.

\section{Geometry scaling}
\label{SI:scaledGeometry}
\begin{figure}[t!]
	\centering
	\includegraphics[width=0.95\columnwidth]{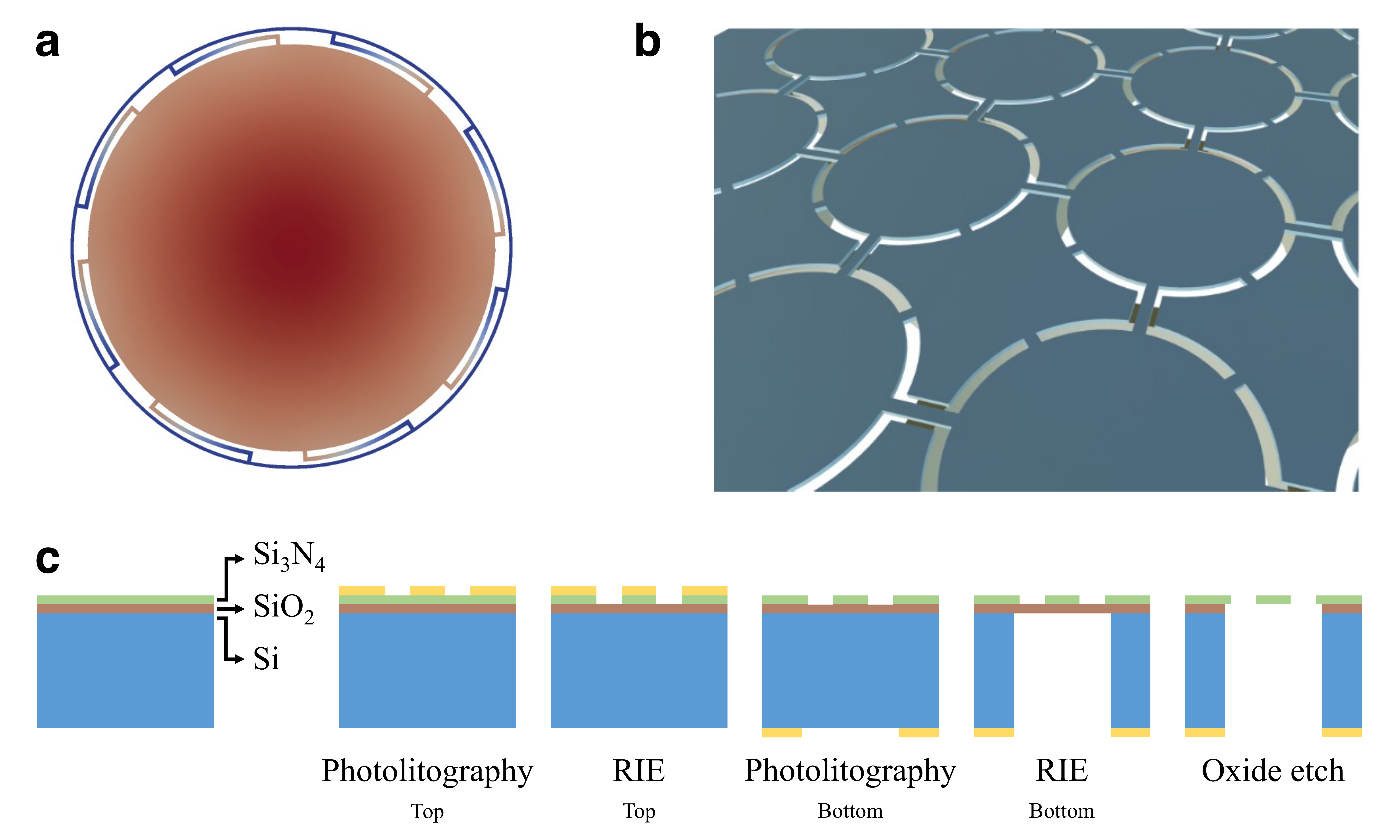}
\caption{Scaling down the geometry to microscopic scales |  
{\fontfamily{phv}\selectfont\textbf{a}}
A $150 \mu m$ diameter drum fabricated on a Silicon-On-Insulator (SOI) wafer can achieve a fundamental frequency of $10.5$ kHz and operate in real time. The second drum mode is located at $16.1 kHz$, providing good spectral separation.
{\fontfamily{phv}\selectfont\textbf{b}}
Proposed approach to interconnect drums in a SOI wafer.
{\fontfamily{phv}\selectfont\textbf{c}}
Fabrication process to produce interconnected drums on a SOI wafer.
}
\label{fig:scaledGeometry}
\end{figure}
In our current experimental platform, reducing the frequency of operation to perform the experiments in real time requires either increasing the lateral dimensions of the design or decreasing the wafer thickness. Increasing the sample size results in dimensions exceeding the 4 inch wafers of our equipment, and decreasing the wafer thickness results in overly brittle devices. However, using multi-layer processes it is possible to scale down the site design to realise microscopic devices that are capable of operating in real time. This can be accomplished by fabricating the resonators on thin films, and using geometric features to reduce the vibrational stiffness. Fig.~\ref{fig:scaledGeometry}a presents a FEM simulation of a $200 nm$ thick drum fabricated on a Silicon-On-Insulator (SOI) wafer. The drum is supported by curved arms that result in a vibrating frequency of $10.5$ kHz. Therefore, the drum can be directly driven with the modulated signal. 

Building the metamaterial on the device layer of a SOI substrate has the effect of decoupling the structural integrity from the elastic interactions between drums. The structural integrity is provided by the substrate, while the resonator sites and the couplings are realised as thin-film resonators on the device layer ~[Fig.~\ref{fig:scaledGeometry}b]. This allows for much thinner structures and therefore lower operating frequencies. The simulated drum has a diameter of $150~\mathrm{\mu m}$, therefore, a $7x7$ lattice can be placed in approximately one square millimeter. Fig.~\ref{fig:scaledGeometry}c shows a fabrication process capable of producing these samples.

\section{Computational methods}
\label{SI:computationalMethods}

The design process described in this work has been implemented primarily in Python. C++ code, parallelised using POSIX threads, is used for the computationally-intensive parts of forward and gradient simulation.

Designing a device involves three different sets of codes. The first set generates the component files for the resonator bodies and links. The component files include the stiffness $K_{\textrm{FEM}}$ and mass $M_{\textrm{FEM}}$ matrices for the components, and metadata to identify degrees of freedom with individual component interfaces. The initial stage of component generation requires MATLAB\textsuperscript{\textregistered} and COMSOL\textsuperscript{\textregistered} licenses. The second set of codes builds the surrogate model used to estimate the dynamic matrices. This code is written in Python 3.7 and can be run using only open source tools. The third group of codes optimises the device geometry, which is exported as an AutoCAD\textsuperscript{\textregistered} script. The sample design part requires access to the Google\textsuperscript{TM} Speech Commands dataset, and can optionally use the Intel\textsuperscript{TM} C++ Compiler instead of the GNU Compiler Collection (GCC). AutoCAD\textsuperscript{\textregistered} is required to convert the optimisation output into a DXF file for fabrication. All codes used to design the sample are provided as supplementary material.

The codes used for generating the component models are in the \textit{ComponentGeneration} folder. The sub-folder \textit{SiteOptimisation} contains the MATLAB script \textit{OptimiseUnitCell.m} optimises the unit cell geometry. The sub-folders \textit{ComsolMatlab/BottomPart}, \textit{ComsolMatlab/CenterPart}, \textit{ComsolMatlab/LeftPart}, \textit{ComsolMatlab/RightPart} and \textit{ComsolMatlab/TopPart} extract the reduced models from COMSOL\textsuperscript{\textregistered} FEM models. Each sub-folder contains two MATLAB scripts, \textit{ExtractMatrices.m}, responsible for extracting the FEM matrices from the COMSOL\textsuperscript{\textregistered} ~file, and \textit{GenerateMatricesRubin.m}, which replaces the internal degrees of freedom by a modal description following the approach by Rubin~\cite{RubinMethod}. The file \textit{ImportCBComponents.ipynb} contains the Python code to sort the degrees of freedom at the interface regions to ensure that full-sample matrices can be directly assembled. 

Codes used for training the surrogate model can be found in the folder \textit{MLModel}. The folder contains three files. \textit{GenerateSites.py} generates pre-computed site components that will be later assembled into the random lattices that are used as training data for the surrogate model. \textit{GenerateRandomLattice\_7x7.py} is used for assembling the pre-computed sites into random lattices and extracting the effective properties. \textit{InterpolateRandomLattice\_7x7.py} is used to train the surrogate model using the mass-spring models generated from the random lattices.

The optimisation codes can be found in the folder \textit{NetworkOptimisation}. The Python file \textit{GenerateTrainingData.ipynb} is used to generate the dataset, by selecting, normalizing, trimming and resampling the original samples from the Google\textsuperscript{TM} Speech Command dataset. The folder \textit{OptimiseSample} contains the code to design a lattice. It requires access to a trained surrogate model to predict lattice properties, plus the set of pre-computed structural components. The optimisation is controlled by the file \textit{PerformOptimisation.py} and accesses the C++ files \textit{prk.cpp} and \textit{grk\_lowmem.cpp} that compute the transmitted amplitudes and gradients respectively. The folder \textit{SecondOrderOptimisation} contains the nonlinear optimization notebook \textit{SpeechRecognitionNonlinear.ipynb} and all associated files used in Figures~\ref{Fig4} and~\ref{fig:nonlinearRealistic}.

%%%%%%%%%%%%%%%%%%%%%%%%%%%%%%%%%%%%%%%%%%%%

\section{Lattice size selection}
\label{SI:latticeSize}

The 7x7 lattice size was chosen because it provides a good balance between optimization complexity and classification precision. We observed that performance degraded by $1.3$ \% when reducing the lattice size to 5x5, while it only improved by $0.5$ \% when increasing the lattice size to 9x9. This was determined by taking the best performance from a batch of 10 simulations per lattice size. For this purpose, the lattice was modelled as a mass-spring model without FEM corrections.

%%%%%%%%%%%%%%%%%%%%%%%%%%%%%%%%%%%%%%%%%%%%

\section{Sound file preparation}
\label{soundFiles}

The sound files composing the test and training set are manually selected from the Google Speech Commands dataset. The list of selected sound files is provided as a supplementary file. For every sample word, we included utterances by a diversity of speakers (>500), genders, and accents. We only excluded samples that were unintelligible to human listeners, truncated, mislabelled or contained significant background noises (intelligible conversations, cars honking, coughing, or electronic hums or buzzes). The files were manually selected based on their hashed file names in a way not traceable to the classification performance of any method or algorithm. The sound files were then trimmed to a duration of $0.6$ s and normalised to a constant mean acoustic power.

%%%%%%%%%%%%%%%%%%%%%%%%%%%%%%%%%%%%%%%%%%%%

\section{Sample fabrication}
\label{SI:fabrication}
\begin{figure}[t!]
\centering
\includegraphics[width=0.95\columnwidth]{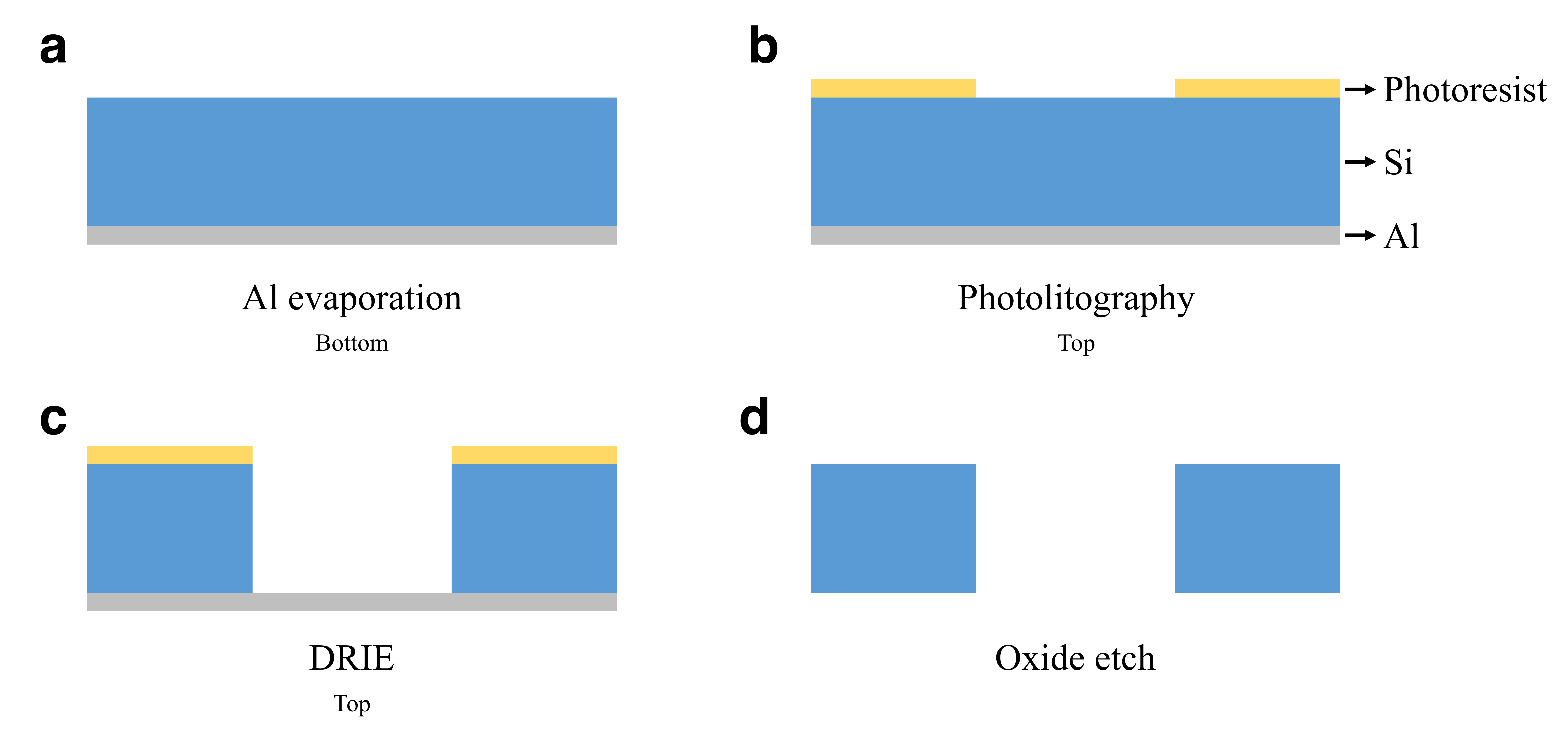}
\caption{Sample fabrication | \textbf{a} The fabrication started with the evaporation of Al on the backside of the wafer. \textbf{b} The wafer was coated with photoresist and the design is patterned. \textbf{c} The silicon was etched with Deep Reactive Ion Etching. \textbf{d} The photoresist and aluminum were removed by wet etching.}
\label{fig:fabrication}
\end{figure}
The sample was fabricated on a double-side polished 100 mm silicon wafer. The wafer thickness is $380$ µm with a measured total thickness variation across the wafer of less than 2 µm. The structures were patterned using standard microfabrication techniques. First, a 5 µm thick layer of Al was deposited on the backside of the wafer using electron beam evaporation~[Fig.~\ref{fig:combi_celloptimisation}a].  The Al layer protected the structure while the silicon was etched. A layer of photoresist with 8 µm thickness was deposited on the top side of the wafer. The sample design was patterned on the photoresist using direct laser writing~[Fig.~\ref{fig:combi_celloptimisation}b]. After hardening the photoresist, the silicon wafer was etched-through using Deep Reactive Ion Etching (DRIE) following a Bosch\textsuperscript{\textregistered}  process alternating etching and passivation cycles~[Fig.~\ref{fig:combi_celloptimisation}c]. After the RIE, the photoresist and aluminum were removed by wet etching~[Fig.~\ref{fig:combi_celloptimisation}d]. 
The sample was clamped between two frames to impose rigid boundary conditions. The frames were obtained following a process similar to the sample fabrication. To bond the frames to the sample, parylene was first deposited on the frames, followed by a bonding process at $350\,\mathrm{{}^{\circ}C}$ . 

%%%%%%%%%%%%%%%%%%%%%%%%%%%%%%%%%%%%%%%%%%%%

\section{Piezoelectric transducer calibration}
\label{SI:piezoCalibration}

Our sample is excited by 28 piezoelectric transducers (Steiner \& Martins Inc, SMD063T07R111) with a thickness-mode resonance at 3 MHz. We calibrate the transducers by measuring the vibration at 56 points along the device boundary while sending Gaussian wavepackets of the form

\begin{equation}
V(t) =e^{\frac{(t-t_0)^2}{2\Delta^2}}\sin\left(2\pi f_c\right) ,
\end{equation}

where $\Delta=15.43\:\mathrm{\mu s}$ is the pulse width and $f_c = 71.28$ kHz is the pulse center frequency. We then Fourier-transform the measured vibration velocity recordings at the boundary points. At every frequency, we define the calibration function by finding the combination of transducer excitation signals that results in the required amplitude at the measured boundary points. Since the number of excitation channels is smaller than the number of measured points, the calibration is done as a least-squares problem using a Moore-Penrose pseudoinverse. To prevent the calibrated excitation amplitude to diverge at higher and lower frequencies, the calibration is only performed between $52$ kHz and $90$ kHz. Beyond this range, the calibration amplitudes and phases are kept constant. Furthermore, during experiments, the output signal is band-passed between $10$ kHz and $110$ kHz to prevent damaging the transducers.